=======================================================================
\documentstyle[12pt,equation]{article}
\epsfverbosetrue

\def\npb#1#2#3{    {\it Nucl. Phys. }{\bf B#1}, #3 (19#2)}
\def\plb#1#2#3{    {\it Phys. Lett. }{\bf B#1}, #3 (19#2)}
\def\prd#1#2#3{    {\it Phys. Rev. }{\bf D#1}, #3 (19#2)}
\def\prep#1#2#3{   {\it Phys. Rep. }{\bf #1}, #3 (19#2)}
\def\prl#1#2#3{    {\it Phys. Rev. Lett. }{\bf #1}, #3 (19#2)}
\def\ptp#1#2#3{    {\it Prog. Theor. Phys. }{\bf #1}, #3 (19#2)}

\def\zpc#1#2#3{    {\it Z. Phys. }{\bf C#1}, #3 (19#2)}

\begin{document}

\newcommand{\ev}{\mbox{events/(kg$\cdot$day)}}
\newcommand{\tanb}{\mbox{$\tan \! \beta$}}
\newcommand{\bsg}{\mbox{$b \rightarrow s \gamma$}}
\newcommand{\br}{\mbox{$Br( b \rightarrow s \gamma) $}}
\newcommand{\om}{\mbox{$\Omega_{\rm LSP} $}}
\newcommand{\omh}{\mbox{$\Omega_{\rm LSP} h^2$}}
\newcommand{\be}{\begin{equation}}
\newcommand{\ee}{\end{equation}}
\newcommand{\een}{\end{subequations}}
\newcommand{\ben}{\begin{subequations}}
\newcommand{\beq}{\begin{eqalignno}}
\newcommand{\eeq}{\end{eqalignno}}

\noindent
\begin{flushright}
KEK--TH--400    \\
KEK Preprint 94--44 \\
DESY 94--096    \\
MAD/PH/835      \\
August 1994
\end{flushright}
\vspace{0.5cm}
\pagestyle{empty}
\begin{center}
{\Large \bf Implications for Supersymmetric Dark Matter Detection from
Radiative $b$ Decays} \\
\vspace{7mm}
{\large Francesca M. Borzumati }                           \\
{\it II.\ Institut f\"ur Theoretische Physik }              \\
{\it Universit\"at Hamburg, 22761 Hamburg, Germany}        \\
\vspace{5mm}
{\large Manuel Drees\footnote{Heisenberg Fellow}}\\
{\it Physics Department, University of Wisconsin, Madison, WI 53706, USA}\\
\vspace{5mm}
{\large Mihoko M. Nojiri\footnote{E--mail: NOJIRIN@JPNKEKVX}}\\
{\it Theory Group, KEK, Oho 1--1, Tsukuba, Ibaraki 305, Japan}
\vspace{7mm}

{\large \bf Abstract}
\end{center}
\begin{quotation}
We point out that combinations of parameters that predict large
counting rates in experiments searching for supersymmetric
dark matter often tend to predict a very large branching ratio for
the inclusive decay \bsg.
%
The recent measurement of this branching ratio,
therefore, indicates that searches for supersymmetric dark matter
might be even more difficult than previously anticipated.
\end{quotation}
\vspace*{1cm}
\clearpage
\noindent
\setcounter{footnote}{0}
\pagestyle{plain}
\setcounter{page}{1}
The Lightest Supersymmetric Particle (LSP) is one of the most
attractive particle physics candidates for the missing dark matter
(DM)\,\cite{book} in the Universe. In the simplest potentially
realistic supersymmetric theory, the Minimal Supersymmetric Standard
Model (MSSM)\,\cite{susyrev}, the LSP is stable by virtue of a
symmetry, the so--called $R-$parity. Calculations\,\cite{relden} have
shown that if the LSP is the lightest of the four neutralino states
present in this model, the relic density of LSPs left over from the
Big Bang is in the desired range over a wide region of the
supersymmetric parameter space. Very broadly this range can be defined
by
\be
0.025 \leq \omh \leq 1,
\label{e1}
\ee
where \om\ is the relic density in units of the closure density, and
$h$ is the Hubble constant in units of 100 km/(sec Mpc). Observations
imply $0.5 \leq h \leq 1$, the lower range, perhaps, being favoured. The
lower bound in (\ref{e1}) then follows from the requirement that there
be enough relic LSPs to form the dark matter haloes of galaxies
($\om \geq 0.1$). The upper bound is equivalent to the constraint that
the Universe be at least 10 billion years old.

Unfortunately relic neutralinos are rather difficult to detect
experimentally.  Here we are interested in direct detection
experiments\,\cite{dirdet}, where one searches for the elastic
scattering of an LSP off a target nucleus. The signal, provided by the
energy deposited in the detector by the recoiling nucleus, has a rate
proportional to the LSP--nucleus scattering cross section.  Partly
because of the Majorana nature of the LSP, this cross section is often
quite small. It can be generally split into two parts\,\cite{gowi}, one
due to spin--spin interactions and the other to scalar
(spin--independent) interactions. For heavy target nuclei the
spin--independent contribution usually dominates the spin dependent
one\,\cite{lspscat}, since it is enhanced by the square of the number
of nucleons in the nucleus in question. This spin--independent
interaction gets contributions from the exchange of the two neutral
scalar Higgs bosons 
of the MSSM as well as from squark exchange. Unless squarks are quite
close in mass to the LSP, the Higgs--exchange contribution usually
dominates. We refer the reader to
refs.\,\cite{lspscat,dn5} for more details on LSP--nucleus
interactions.

Thus, the LSP--nucleus scattering cross section depends, in general,
on many parameters: the gaugino mass $M_2$, the higgsino mass $\mu$
and ratio of Higgs vacuum expectation values \tanb\ entering the
neutralino mass matrix\footnote{ We assume the usual unification
 relation between the $U(1)$ gaugino mass $M_1$ and the $SU(2)$
 gaugino mass $M_2$, $M_1 = (5/3) \tan^2 \theta_W M_2 \simeq 0.5
 M_2$.}\,\cite{susyrev};
the squark masses and mixings; and the masses and couplings of the
Higgs bosons. At the tree level the Higgs sector of the
MSSM\,\cite{guha1} is completely specified in terms of two parameters,
which we take to be \tanb\ and the mass $m_P$ of the pseudoscalar
Higgs boson. As it is well known, radiative corrections\,\cite{higcor}
to the mass of scalar Higgs bosons introduce also a dependence on the
mass of the top-quark, $m_t$, as well as on the parameters describing
the mass matrix for the scalar superpartners of the top--quark, or
stop $\tilde{t}$ (see below). In our analysis we include these
corrections using the effective potential
method\,\cite{effpot}\footnote{ Since not only corrections growing as
 $\log ({m_{\tilde t}}/{m_t})$ are included, it is technically easier
 to present our results for fixed $m_P$, rather than for fixed mass of
 one of the scalar Higgs bosons.}.
As for the slepton masses, needed for the calculation of the LSP relic
density, we follow the conventional choice made in DM searches: we
assume that the squared masses of all sfermions get the same soft
supersymmetry breaking contribution $m^2$ along the diagonal of their
respective mass matrices. Our main result is independent of this
assumption. Finally, the specification of the neutralino mass matrix,
due to gauge invariance, completely determines also the chargino
sector.

Having fixed the (s)particle spectrum it is imperative to first check
for consistency with experimental and theoretical constraints before
we use this spectrum to predict LSP detection rates. In particular,
$M_2, \ \mu$ and \tanb\ must be chosen such that charginos and
neutralinos escape detection at LEP\,\cite{lepino}.  Similarly,
searches for neutral Higgs bosons at LEP\,\cite{lepino} constrain the
parameters of the Higgs sector.

There is yet another constraint which has so far been ignored in
estimates of LSP detection rates. The CLEO II collaboration has
measured\,\cite{cleo} the branching ratio for inclusive \bsg\ decays to be
\be \label{en1}
\br = (2.32 \pm 0.51 \pm 0.29 \pm 0.32) \cdot 10^{-4},
\ee
where the errors are statistical, experimental systematics and theoretical
systematics (due to the extrapolation from the observed part of the photon
spectrum), respectively. Adding all errors in quadrature, this implies 95\%
c.l. upper and lower limits on this branching ratio of $3.4 \cdot 10^{-4}$ and
$1.2 \cdot 10^{-4}$, respectively. These bounds are relevant for LSP searches
since within the MSSM the \br\ is determined\,\cite{bsmssm} by the {\em same}
parameters that determine LSP detection rates, i.e. the masses and mixings of
squarks and charginos as well as the mass of the charged Higgs boson,
$m_{H^\pm}$. This is related to $m_P$ by\footnote{  Eq.(\ref{e2}) holds at
tree level. Radiative corrections to this  relation are very
small\,\cite{effpot} unless one somewhat  artificially allows $\tanb < 1$.}:
\be
m^2_{H^\pm} = m_P^2 + m_W^2,
\label{e2}
\ee
where $m_W \simeq 80$\,GeV is the mass of the $W$ bosons. In
particular, a light charged Higgs gives a large positive contribution
to the amplitude ${\cal A}(\bsg)$.  Loops involving charginos (or, in
general, gluinos) and squarks, in contrast, give contributions with
either sign and decouple in the limit of large sparticle
masses. Therefore, spectra of supersymmetric particles with rather
light Higgses when sparticles are taken to be heavy, although well
suited for DM detection, tend to give results for the \br\ similar to
the ones obtained in the two--Higgs--doublet model
(type~II)\,\cite{bshig1,bshig2}. Thus, one may expect clashes with the
experimental upper bound on this decay in regions of parameter space
where the counting rates are at the highest values.

In order to quantify this statement we have to specify the amount of
flavor mixing in the squark sector through which transitions from
the third to the second generation of quarks, such as the
decay \bsg, can occur.  As mentioned, mimicking as closely as possible the
treatment of squark masses in previous analyses of LSP
detection\,\cite{lspscat,lspdet}, we assume that all sfermions have
the same soft supersymmetry breaking mass. This implies that
{\em no} contributions to the decay \bsg\ can come from loops mediated
by neutral gauginos, gluinos or neutralinos. Flavor mixing in the
quark sector, however, will introduce some mixing in the squark sector
as well.  Following ref.\,\cite{bsmssm}, we work in a quark basis
in which current and mass eigenstates coincide for right handed
quarks as well as left--handed down--type quarks.  Flavor mixing,
therefore, affects only left--handed $u-$type squarks, $\tilde{u}$
in the $6 \times 6$ mass matrix:
\be
{\cal M}^2_{\tilde u} = \mbox{$ \left(
       \begin{array}{cc}
{\cal M}^2_{\tilde{u}_L} & {\cal M}^2_{\tilde{u}_{LR}} \\
\left( {\cal M}^2_{\tilde{u}_{LR}} \right)^{\dag}
                         & {\cal M}^2_{\tilde{u}_R}
       \end{array} \right)$}.
\label{e3}
\ee
The $3 \times 3$ left--left, right--right and left--right
mixing submatrices
${\cal M}^2_{\tilde{u}_L}, \ {\cal M}^2_{\tilde{u}_{LR}}$ and
${\cal M}^2_{\tilde{u}_R}$ are given by:
\ben
\label{e4}
\beq
\left( {\cal M}^2_{\tilde{u}_L} \right)_{ij} &
     = (m^2 + 0.35 m_Z^2 \cos \! 2 \beta)\, \delta_{ij}
            + m_t^2 V_{3i}^{\ast} V_{3j} ;
 \label{e4a} \\
\left( {\cal M}^2_{\tilde{u}_R} \right)_{ij} &
     = (m^2 + 0.15 m_Z^2 \cos \! 2 \beta)\, \delta_{ij} +
         m_t^2 \delta_{i3} \delta_{j3};
 \label{e4b} \\
\left( {\cal M}^2_{\tilde{u}_{LR}} \right)_{ij} &
     = - (A_t + \mu \cot \! \beta)\, m_t V_{3i}^{\ast} \delta_{j3}.
 \label{e4c}
\eeq
\een
The symbols $V_{ij}$ indicate here elements of the CKM mixing matrix,
$\mu$ is the mass parameter entering the neutralino mass matrix, and
$A_t$ is a soft supersymmetry breaking parameter of order $m$.  When
writing eqs.\,(\ref{e4}) we have neglected all Yukawa couplings except
for the top quark.  Similarly, the left--right mixing in
(\ref{e4c}) is significant only for the third generation of squarks.

We are now in a position to discuss quantitatively the correlation
between the relic LSP detection rate and the \br. For definiteness we
focus on a detector consisting of isotopically pure $^{76}$Ge since
such a device is now under construction. The impact of the measurement
of \br\ on the prospects for direct relic LSP detection is very
similar for detectors of different materials as long as the total
LSP--nucleus cross section is dominated by spin--independent
interactions, which is true in almost all cases.  The next round of
experiments is expected to reach a sensitivity of about 0.1 \ev\ which
improves on the current best limits
\cite{lspsearch} by about a factor of 100.

We show in figs.\,1 the LSP counting rate in such a detector
(solid lines) as well as the branching ratio for \bsg\ (dashed
lines) as function of various parameters of the MSSM and for fixed
top--quark mass, $m_t=175$\,GeV.  We give results for the case of a
heavy LSP ($M_2 = 500$\,GeV and $\mu=400$\,GeV, giving
$m_{\rm LSP}\simeq$ 200\,GeV) and the case of a much lighter one
($M_2 =100$\,GeV, $\mu=-100$\,GeV, giving
$m_{\rm LSP} \simeq 50 $\,GeV).  We fix the remaining supersymmetric
parameters to be, in general, $A_t=0$, $\tanb=2$, $m_P = 150$\,GeV,
and choose $m$ to be respectively $m=500$\,GeV and $m=200$\,GeV
in the case of the
heavy and light LSP. We then deviate from these points in the
supersymmetric parameter space by varying $m_P$ (fig.\,1a), \tanb\
(1b), $m$ (1c) or $A_t$ (1d).

We have chosen $\vert \mu \vert > M_1$ so that
the LSP is predominantly a gaugino;
this is necessary \cite{relden} to satisfy the lower bound on \omh\ in
(\ref{e1}).  In both cases, however, the higgsino components of the
LSP are still quite substantial, leading to sizable couplings of the
LSP to Higgs bosons. The LSP detection rate can therefore be quite
large {\em if} Higgs bosons are light.

This is illustrated in fig.\,1a, where the dependence on $m_P$ is
shown. If $M_2, \ m$ and $\mu$ are large (upper pair of curves), $m_P$
can be chosen as small as 50\,GeV without violating direct Higgs
search limits. This would lead to an LSP counting rate of 0.2 \ev,
in principle observable in the next round of direct detection
experiments.  Nevertheless, demanding that $\br $ be $\leq 3.4 \cdot
10^{-4}$ implies $m_P \geq 500$\,GeV and a counting rate of less than
0.01 \ev, a value too small to be detectable in the near
future. Notice, however, that no error in the theoretical
prediction for $\br$ has been assumed, as yet. This
discussion is postponed to a later point of this paper.

For our second choice of parameters (lower pair of curves) the lower
bound on $m_P$ is set by Higgs searches. In this case, which is
characterized by rather small squark and chargino masses, there is a
sizable contribution to the decay \bsg\ from chargino--squark loops
which interferes {\em destructively} with the contributions from $W$
and $H^{\pm}$ loops. The Higgs contributions decouple in the limit of
large $m_P$. As a result this scenario gives for $m_P > 350$\,GeV,
values of \br\ {\em below} the Standard Model (SM)
prediction of $2.9 \cdot 10^{-4}$\,\cite{bssm}. Moreover, the lower bound
on $m_P$ is relaxed to 250 GeV in this scenario. It
should be noted that for fixed $m_P$ this scenario gives a
significantly smaller counting rate than the case with heavy LSP,
even though a lighter LSP means a larger LSP flux (the mass
density of relic neutralinos in the vicinity of the solar system is
assumed to be fixed) and less suppression due to nuclear form factors
\cite{lspdet}. The reason is that negative values of $\mu$
(in the convention of ref.\,\cite{guha1}, which we follow throughout)
always imply less gaugino--higgsino mixing in the neutralino sector,
and hence smaller couplings of the LSP to Higgs bosons. As a result the
lower bound on $m_P$ corresponds to a maximal LSP counting rate as small
as 0.007 \ev\ for this case.

In fig.\,1b we show the \tanb\ dependence for our two choices of the LSP mass.
Here we have chosen $\mu=-300$\,GeV in the light LSP case, in order to ensure
$\omh \geq 0.025$ for a sizable range of \tanb. We have terminated the curve
for $M_2=100$\,GeV at $\tanb=33$ since larger values give a too small LSP
relic density. The lower bound on \tanb\ is given by the requirement that the
Higgs boson escapes detection at LEP\,\cite{lepino}.  We see that the \tanb\
dependence of the LSP counting rate is quite similar in both cases. At first,
the rate decreases with increasing \tanb\ since the mass $m_{h^0}$ of the
light neutral Higgs $h^0$ increases. This mass, however, remains essentially
constant for $\tanb \geq 5$. At the same time, the coupling of the heavier
neutral Higgs $H^0$ to down--type quarks increases roughly as $\tanb$. In the
case of the light LSP, the coupling of the LSP to $H^0$ is less suppressed
than the coupling to $h^0$. In addition, in this scenario, squark exchange
contributions are not entirely negligible (recall that it is for
$m=200$\,GeV). As a result, the counting rate grows faster with \tanb\ in the
case of the light LSP than in the case of the heavy one.

The difference between the two cases is much more dramatic for the
\br. The contribution from $H^{\pm}$ loops becomes independent of
\tanb\ roughly for $\tanb > 3$. In the scenario with heavy LSP and
even heavier charginos and squarks, the contribution from sparticle
loops is always small compared to the SM and Higgs contributions,
leading to an overall very weak dependence of the branching ratio on
\tanb. Notice that the entire curve lies well above the experimental upper
bound of $3.4 \cdot 10^{-4}$. In contrast, in the case with a light
LSP and comparatively light charginos and squarks, sparticle loops do
contribute significantly. The observed strong \tanb\ dependence of
this contribution is due to the fact that the
chargino--$b$--$\tilde{t}$
interaction contains \cite{guha1,bsmssm} a term proportional to the
bottom Yukawa coupling, which grows as $ \tanb$ for large \tanb. For
the given values of $M_2$ and $m$ but positive $\mu$ the contribution
from this term would be positive, leading to predictions for \br\
rapidly growing and quickly exceeding the experimental upper
bound. For the given case of negative $\mu$ this contribution
interferes destructively with the $W$ and $H^{\pm}$ loops; for $\tanb > 24$
one gets into conflict with the experimental lower bound
which implies a counting rate of less than 0.2 \ev\ in the case with
light LSP.

The dependence on the parameter $m$ is shown in fig.\,1c. The LSP
counting rate falls monotonically with increasing $m$ since, due to
the above mentioned radiative corrections to the Higgs
sector\,\cite{higcor,effpot}, $m_{h^0}$ increases with $m$. In the
light LSP scenario the lower bound on $m$ comes from the Higgs search
limits, whereas in the heavy LSP case this bound is set by the
requirement that the lightest squark (essentially a
$\tilde{t}$--squark) be heavier than the lightest neutralino. A
charged LSP, in
fact, would result in too large an abundance of exotic isotopes
\cite{iso} such as the one with a squark bound to a 
nucleus. At the lower end of $m$ squark exchange contributions to
LSP--nucleus scattering are quite important, including the
${\cal O}(m^{-4}_{\tilde q})$ terms discussed in ref.\cite{dn5}; this
explains the rapid decrease of the expected counting rate with
increasing $m$ in this region.

In the two cases of light and heavy LSP, the \br\ increases with
increasing $m$, since in both cases chargino contributions interfere
destructively with $W$ and $H^{\pm}$ contributions. The $m$ dependence
is much stronger for the light LSP due to the presence of lighter
charginos and hence potentially larger contributions from sparticle
loops. In the limit of large squark masses these contributions are
always very small and practically independent of the chargino mass,
leading to the observed convergence of the two curves at the higher
end of $m$. Note that the entire curve for the heavy LSP is once
again above the experimental upper bound on the branching ratio, while in
the case of light LSP only the region close to the lower bound on $m$
imposed
by Higgs searches is marginally compatible with the upper bound on
the branching ratio.

In fig.\,1d we show the dependence on the parameter
$A \equiv A_t/m$. Compared to the previous three figures we observe a
rather mild variation of the LSP counting rate. The effect is almost
entirely due to the radiative corrections to $m_{h^0}$, which depend
on $A_t$ in a nontrivial way \cite{higcor,effpot}. In particular,
$m_{h^0}$ reaches a minimum
when the combination
$A_t + \mu \cot \!\beta$ of eq.\,(\ref{e4c}) is very small. Increasing
it, at first increases $m_{h^0}$  which then reaches a maximum at some
finite value of $A$. Increasing the combination
$|A_t + \mu \cot \!  \beta|$ even further, reduces $m_{h^0}$, as it
is very visibly shown by the curve relative to the heavy LSP case.
The upper bound on $A$, at which both sets of curves are stopped, is
determined by the requirement that the lightest squark be heavier
than the lightest neutralino. The observed variation of the counting
rate then follows from the fact that the $h^0$--exchange contribution
to the LSP--nucleus cross section scales like the {\it inverse} of
$m_{h^0}^4$.

\setcounter{footnote}{0}
In contrast, the \br\ increases or decreases monotonically with
increasing $A$. Once again only the contribution from chargino--squark
loops changes when $A$ is varied. The absolute value of this
contribution reaches a minimum at small values of
$|A_t + \mu \cot \! \beta|$
where the lightest squark mass is maximal (recall that
off--diagonal entries in the squark mass matrix tend to reduce the
smallest eigenvalue and increase the largest one). We observe that the
sign of this contribution is flipped when going from positive to
negative $A$, since the sign of the left--right mixing terms changes
for the set of supersymmetric parameters chosen here. The sign of this
contribution depends also on the sign of $\mu$, and its absolute size
is larger for the case with light LSP.  Hence, the slope of the curve
for the light LSP scenario, for which we have taken $\mu < 0$, is
opposite in sign and larger in magnitude than the slope for the curve
relative to the heavy LSP and positive $\mu$. Notice that this latter
scenario again violates the upper bound on the branching ratio over the entire
parameter range shown here, while in the case of a light
LSP this bound is violated for $A > -0.7$.

Finally, the dependence on $M_2$ and $\mu$ is shown in fig.\,2, where
we have fixed $A_t = 0$, $\tanb = 2$, $ m_P$ = 200\,GeV and
$m_t = 175$\,GeV and we have chosen
$m = \min(120\,{\rm GeV}$, $2 m_{\rm LSP})$.
The shaded regions in both frames are excluded by LEP
searches for charginos, neutralinos and Higgs bosons \cite{lepino}
(region of small $|\mu|$ or small $M_2$), or by the requirement that
the lightest squark (which again is mostly a $\tilde{t}$--squark) is
heavier than the lightest neutralino (regions of large $|\mu|$ and
small or moderate $M_2$).  The short--dashed lines
indicate contours of constant $\br = 3.4 \cdot 10^{-4}$: larger values
of $\br$ are obtained above these lines and smaller below.  We also
show contours of constant LSP counting rate $ = 1, \ 0.1$, and 0.01
\ev\ for $\mu > 0$, and 0.01 and 0.003 \ev\ for $\mu < 0$ (solid lines).
The dotted lines are contours of constant $\omh = 0.025$\footnote{To
 avoid figures too cluttered, we have
omitted very narrow regions with $\omh < 0.025$ where $s-$channel exchange
diagrams become resonant.}. 
In the regions of small $|\mu|$ and large $M_2$ the LSP is higgsino--like and
its relic density is too small to be of cosmological interest \cite{relden}.

We observe the well--known \cite{lspdet} correlation between large LSP
counting rate and small LSP relic density; in particular, about 50\% of
the region where the counting rate exceeds 0.1 \ev\ (for fixed
neutralino flux!) lies within the region with $\omh < 0.025$
(lower frame).
It has been suggested in the literature to re--scale the
counting rate in such regions in order to take into account the
reduced LSP flux. We prefer to discard these regions altogether, since
here the LSP can only make an almost negligible contribution to the
solution of the DM problem\footnote{ Recall that there is now rather
 solid evidence \cite{tutalk} that $\Omega > 0.1$ on bigger than
 galactic length scales.}.

Prospects for direct LSP detection become even less promising once we
require the \br\ to be below its experimental upper bound. Only the
little region at small $M_2$ and $\mu \simeq 450$\,GeV survives for
positive $\mu$, while for $\mu < 0$ the somewhat larger region to the
right and below the short--dashed curve remains acceptable.
The whole region where the counting rate exceeds 0.1 \ev\
is now excluded. For the allowed region with
positive $\mu$ the counting rate is even below 0.01 \ev.
The implementation of the experimental bound on
$\br$ implies that, for the values of supersymmetric
parameters chosen here, the maximal LSP counting rate in $^{76}$Ge
is about 0.007 and 0.02 \ev\ for positive and
negative $\mu$, respectively. Notice that
a wider portion of parameter space survives for $\mu < 0$
where the LSP counting rate is usually smaller.
In this region, in fact, the expected \br\ gets destructively
interfering contributions from chargino--squark
loops (at least in the region of small or moderate $M_2$). 

At this point we should warn the reader that our predictions for both
the LSP counting rate and the branching ratio of radiative $b$ decays
are fraught with substantial theoretical uncertainties. The LSP
counting rate obviously depends on the local density and velocity
distribution of relic neutralinos. In our calculation we have used the
standard values \cite{tu} of 0.3\,GeV/cm$^3$ for the LSP mass density
and 320 km/sec for their velocity dispersion. The calculation of the
LSP--nucleus scattering cross section also suffers from uncertainties,
the most important one being the value of the nucleonic matrix element
$\langle p| m_s \bar{s} s | p \rangle$, which we have taken to be
130\,MeV\,\cite{strmat}. Varying this value within the range favoured by
model calculations can change the prediction for the LSP counting rate
by as much as a factor of 2.

The uncertainty in our prediction for \br\ stems primarily from the
fact that the present calculation is in some sense still at the
leading order in perturbative QCD. As a result there is a rather
strong dependence on the value of the renormalization scale $Q_0$ that
is used in the calculation. The resulting uncertainty has been
emphasized by Ali and Greub\,\cite{AG}
and has more recently been elaborated by
Buras et al. \cite{bssm}, who have also included uncertainties due to
the experimental errors on parameters entering the prediction of this
branching ratio in their analysis. Within the SM they find an overall
theoretical ``1 $\sigma$" uncertainty of about $\pm 25 \%$, which
includes the uncertainty that results from varying $Q_0$ from 2.5 to
10\,GeV (i.e.  approximately from $m_b/2$ to $2 m_b$). We have
(linearly) added an additional 8\% uncertainty, which is the size of
an already known part of higher order QCD corrections \cite{bssm}.
Although strictly speaking a statistical meaning cannot
be assigned to the theoretical uncertainty, nevertheless, for the time
being, one can obtain a conservative estimate of the branching
ratio by allowing a ``1 $\sigma$ downward fluctuation" due to this
theoretical uncertainty.\footnote{Very recently another theoretical estimate
of \br\ has appeared \cite{ciuc}, where also parts of the next--to--leading
order contributions to the relevant matrix of anomalous dimensions have been
included. This introduces a strong renormalization
scheme dependence. We prefer to follow here
the approach of ref.\cite{bssm} where
these terms are not included. The result of ref.\cite{ciuc} falls within our
theoretical error band.} We should mention here that the {\em
relative} theoretical uncertainty is often smaller in the MSSM than in
the standard model. The reason is that a purely QCD--induced additive
contribution to the \bsg\ matrix element, which contributes greatly to
the QCD uncertainty, becomes less important when additional terms are
added with the same sign as the $W$--loop contribution present in the
SM.

Contours where we take the value $3.4 \cdot 10^{-4}$ as this
conservative (low) estimate of the $\br$ are shown by the long--short dashed
lines in fig.\,2. We see that, for the given set of parameters, most
of the $(M_2,\mu)$ half--plane with $\mu>0$ is still excluded even if this
lower theoretical estimate
of the branching ratio is indeed correct, while for $\mu < 0$ the allowed
region grows substantially. If this lower theoretical estimate is used, the
maximal counting rate for $\mu>0$ increases to about 0.02 \ev; for $\mu<0$ the
upper bound on the counting rate is mostly due to direct experimental SUSY
searches, but it is noteworthy that now the entire region of this half--plane
where the counting rate exceeds 0.01 \ev\ is allowed.

In order to give the reader a feeling of
how far above the experimental upper
bound most the half--plane with positive $\mu$ lies, we also show a contour
where the lower theoretical estimate yields $\br = 4.2 \cdot 10^{-4}$
(long--short--dashed curve), which is the 95\% c.l. upper limit if
all errors in
eq.(\ref{en1}) are added linearly. Only in the region to the left and below
this line, as well as in the region where both $M_2$ and $\mu$ are very large
so that sparticles decouple, does the lower estimate lie above this value.
This means that most of the region with $\mu>0$ is allowed only if we
simultaneously assume a large downward fluctuation of the measured \br\
(including systematic errors) {\em and} a theoretical estimate for the
branching ratio at the lower edge of the expected range. The contour where
our {\em central} estimate for $\br = 4.2 \cdot 10^{-4}$ almost coincides
with the contour where the lower estimate yields $3.4 \cdot 10^{-4}$, which
would again exclude most of the positive half--plane for the present choice of
$m_P, \ \tanb, m$ and $A$.

If the lower theoretical estimate for \br\ turns out to be correct the
interpretation of figs. 1 would also change somewhat. For example, for the
heavy LSP case shown in fig.\,1a a bound $m_P \geq 300$\,GeV would
result, leading to a reduction of the maximal counting rate by a
factor of 20. The heavy LSP case in figs.\,1b--d would still be excluded over
the whole range of parameters shown, however.
Finally, it should be clear from the above discussion
that our quantitative results depend on the ansatz for the
squark mass matrices. It is conceivable that {\em ad hoc}
modifications of this simple ansatz would allow to partially
circumvent the constraints imposed by the experimental bounds on
\br. It remains still true, however, that in general one is
likely to get into conflict with these bounds by simultaneously
choosing a heavy sparticle spectrum and light Higgs bosons.

In conclusion, we have pointed out that the experimental upper bound
\cite{cleo} on the branching ratio for inclusive \bsg\ decays imposes
somewhat strong constraints on the region of parameter space where
sizable counting rates for relic neutralinos are expected. In some
cases, the lower bound on the branching ratio is also relevant. In
spite of considerable theoretical uncertainties in the estimates of both
the LSP counting rate and the \br, it seems fair to say that prospects
for the next round of LSP detection experiments look much bleaker
once the CLEO constraint is incorporated in the analysis. The main
reason for this is that both the counting rate and the branching
ratio increase with decreasing mass of the Higgs bosons in the theory;
the experimental upper limit on the latter hence reduces the maximal
possible value of the former.

In this paper we focussed on direct LSP detection experiments. The
expected signal rate in experiments looking for LSP annihilation in
the center of the Earth or Sun\,\cite{annexp} is also
proportional to
LSP--nucleus scattering cross sections. The upper limit on the \br\
therefore will tend to reduce the maximal signal that can be
expected in such experiments as well.

Note that we have assumed in our analysis that sparticle masses,
$\mu$, and Higgs masses can all be varied independently from each
other; the same assumption has been made in almost all previous
studies of LSP detection. Given that the main motivation for the
introduction of weak scale supersymmetry is to help understanding
electroweak symmetry breaking, such an assumption appears quite
unnatural. One would rather expect these masses to be of roughly the
same size. This statement can be quantified in so--called minimal
supergravity theories \cite{sugra}, where the mechanism of radiative
gauge symmetry breaking ensures that sparticles masses, $m_P$ and
$\mu$ are strongly correlated. In such models the upper bound on \br\
is more easily satisfied 
\cite{bsmssm,bssugra}. The price one has to pay, though, is that
the expected LSP
counting rates are usually very low \cite{dn5}, of order $10^{-3}$
\ev\ or even less in a $^{76}$Ge detector. The main result of this
paper is that now a purely {\em experimental} constraint seems to
force us closer to regions of parameter space favoured by these
supergravity models, which are at the same time theoretically very
appealing but difficult to probe by LSP detection experiments.

\subsection*{Acknowledgements}
M.\,D. acknowledges the hospitality of theory group at
DESY, where part of this work was carried out. His work was supported
in part by the U.S. Department of Energy under contract
No. DE-AC02-76ER00881, by the Wisconsin Research Committee with funds
granted by the Wisconsin Alumni Research Foundation, as well as by a
grant from the Deutsche Forschungsgemeinschaft under the Heisenberg
program.  F.M.\,B.'s work was supported by grant no. 05 5HH 91P(8)
from the Bundesministerium for Forschung und Technologie, Bonn,
Germany. M.M.\,N. was supported by grant no. 06740326 from the
Japanese Ministry of Education, Science and Culture.

\newpage
\section*{Figure Captions}
\renewcommand{\labelenumi}{Fig.\arabic{enumi}}
\begin{enumerate}
\item    
 Dependence of the LSP counting rate in a $^{76}$Ge
 detector (solid lines) and \br\ (dashed lines) on: the mass $m_P$ of
 the pseudoscalar Higgs boson\,(a), the ratio of vacuum expectation
 values \tanb\,(b), the scalar mass parameter $m$\,(c) and the soft
 breaking parameter $A \!\equiv \!A_t/m$\,(d). Results are presented
 for $m_t \!= \!175$\,GeV and for the case of a heavy
 ($M_2\!=\!500$\,GeV) and a light ($M_2\!=\!100$ GeV) LSP. The
 remaining MSSM parameters are specified in the text.

\item   
Contours of LSP counting rate in a
$^{76}$Ge detector (solid lines) equal to
1\,0.1\,0.01 \ev\,(for positive $\mu$), 0.01,\,003 \ev\,(for
negative $\mu$)
and of constant $\br\!=\!3.4 \cdot 10^{-4}$
(dashed and dot--dashed lines) in the $(M_2,\mu)$ plane. The
values of the other parameters are $m_t\!=\!175$\,GeV, $A_t\!=\!0$,
$\tanb\!=\!2$, $m_P\!=\!200$\,GeV and $m\!=\!\min(2m_{\rm LSP},
120$\,GeV). In each frame the shaded regions are
excluded by sparticle and Higgs searches and by the requirement
that the LSP be neutral; the regions enclosed by the dotted
lines have a very small LSP relic density, $\omh \!< \!0.025$. The
short--dashed contours of the \br\ have been computed using
$Q_0=5\,$GeV; the dot--dashed ones allow for
some theoretical uncertainty, as described in the text, and the
long--short--dashed contour correspond to
$\br\!=\!4.2 \cdot 10^{-4}$.

\end{enumerate}

\end{document}